\documentclass[a4paper,twoside,notitlepage,fleqn]{report}
\usepackage{lb_rep}
\usepackage{graphicx}
\usepackage{ifthen}
\usepackage{longtable}
\usepackage{epsfig}
\usepackage{amssymb}
\usepackage{rotating}
\usepackage{lscape}
\usepackage[small,bf,nooneline]{caption2}
\renewcommand{\captionlabeldelim}{.}

\begin{document}

\mbox{}
\vspace{3cm}

\noindent
{\bf\Huge UV and EUV Instruments}
\\ \\ \\
January 2010
\\ \\
to appear in:\\
{\bf Landolt-B\"ornstein, New Series VI/4A, Astronomy, Astrophysics, and Cosmology;
Instruments and Methods, ed. J.E. Tr\"umper,
Springer-Verlag, Berlin, 2010}
\\ \\ \\
{\bf K. Werner}\\
Institut for Astronomy and Astrophysics\\
Kepler Center for Astro and Particle Physics\\
University of T\"ubingen\\
Sand 1, 72076 T\"ubingen, Germany

\newpage

\section{UV and EUV instruments}
\label{euvinstruments} 

We describe telescopes and instruments that were developed and used for
astronomical research in the ultraviolet (UV) and extreme ultraviolet (EUV)
regions of the electromagnetic spectrum. The wavelength ranges covered by these
bands are not uniquely defined. We use the following convention here: The EUV
and UV span the regions $\lambda\lambda$~100--912\,\AA\ and 912--3000\,\AA,
respectively. The limitation between both ranges is a natural choice, because
the hydrogen Lyman absorption edge is located at $\lambda$~912\,\AA. At smaller
wavelengths, astronomical sources are strongly absorbed by the interstellar
medium. It also marks a technical limit, because telescopes and instruments are
of different design. In the EUV range, the technology is strongly related to
that utilized in X-ray astronomy, while in the UV range the instruments in many
cases have their roots in optical astronomy. We will, therefore, describe the UV
and EUV instruments in appropriate conciseness and refer to the respective
chapters of this volume for more technical details.

The $\lambda$~100\,\AA\ limit of the EUV range to the soft X-ray range is fuzzy,
because it is covered by both, specific X-ray and EUV instrumentation, and
because there is no scientific motivation for a sharp distinction between these
regions. As a consequence, there will be some minor overlap with the X-ray
chapter of this volume concerning the presentation of specific instruments.  In
contrast, the long-wavelength limit of the UV range to the optical region is an
obvious choice. The $\lambda>3000$\,\AA\ region is accessible with ground-based
telescopes, whereas investigation shortwards of this limit requires observations
from above Earth's atmosphere.

All facts presented here are mainly taken from original papers referred to in
Table~\ref{table1a}. A number of reviews are useful sources, e.g.: [69Wil] and
[72Ble] on UV astronomy, [71Car] on electronic imaging devices, [91Bob] on the
cosmic far-UV ($\lambda\lambda$~1000--2000\,\AA) background, and [00Bow] on EUV
astronomy. Significant books on the state-of-the-art in the relatively concise
field of EUV astronomy (in comparison to the UV) are two conference proceedings
[91Mal], [96Bow], and a very useful, exhaustive presentation by [03Bar].

We restrict our description here to space-based instruments, i.e., to those
facilities that were carried by satellites or by interplanetary and lunar
probes. For conciseness, we do not address the issue of sounding rocket
experiments, although they have been essential for the development of UV and EUV
instruments and obtaining early scientific results. Likewise, we will not
describe stratospheric balloon experiments, which exploit the fact that Earth's
atmosphere beyond a height of $\approx$40\,km above sea level is sufficiently
transparent in the $\lambda\lambda$~2000--3000\,\AA\ region to perform
astronomical observations.

For UV and EUV instrumentation used to observe the Sun, we refer to another
chapter of this volume. Solar observatories (e.g. SOHO, STEREO) occasionally performed stellar
observations for calibration purposes [07Val]. We also disregard here UV/EUV instruments that were used
by space probes to observe exclusively solar-system objects (Moon, planets,
interplanetary matter, etc.). But we will point out UV/EUV instruments of such
probes that were also used to perform astronomical observations of Galactic and
extragalactic objects.

In the following, we describe EUV and UV instruments in two separate
sections. In the sense just outlined, Table~\ref{table1a} represents a complete
compilation of all EUV and UV instruments that flew on space probes, in
chronological order of their launch date. For details on the instrumentation
references are made to Table~\ref{tabinstr}.

Most instruments were carried on \emph{Earth-orbiting satellites}, the first few
of them were unstabilized. Important measurements were contributed from
instruments that flew on missions to planets. During \emph{interplanetary
cruise}  phases, astronomical observations were performed with instruments
primarily devoted to the study of  planetary atmospheres: Venera 2 \& 3 to
Venus, Mariner 9 to Mars (and in Mars orbit), and Voyager 1 \& 2 to the outer
solar system. Other instruments were operated on \emph{manned space vehicles:}
Gemini (during extravehicular stand-up activities by astronauts), Apollo (on
the lunar surface and during translunar coast), Soyuz, the MIR and Skylab space
stations, and quite frequently on several Space Shuttle flights.

\subsection{EUV instruments}

An excellent summary about the current state of technology for EUV
space-instrumentation is given in [03Bar].

\subsubsection{EUV detectors}

EUV detector developments were determined by two approaches (photon counting
devices):

\begin{itemize}
\item Proportional counters (as used in X-ray astronomy), and
\item Photomultipliers (as used in UV range).
\end{itemize}

The problem with proportional counters is, that detector windows should not
absorb photons (this is a much smaller problem in the X-ray range). Therefore,
thinner windows or plastic windows are used. But significant gas losses occur,
hence, ``thin window, gasflow'' proportional counters were developed. But these
are limited to $\lambda<200$\,\AA,  therefore other approaches are more common
in EUV astronomy: CEM (channel electron multiplier) and MCP (microchannel
plates). The CEM is a compact, windowless version of a
photomultiplier. Therefore, photons at all EUV wavelengths can be detected.

Two-dimensional (2-d) position-sensitive detectors are required for observations with focussing
instruments and for an efficient recording of spectra.  Three techniques have
been advanced:

\begin{itemize}

\item MCP detectors (e.g. used in Chandra/LETG, EUVE, ROSAT/WFC, EXOSAT/CMA,
Einstein/HRI). Essentially, MCPs  are arrays of single CEMs. They have no energy
resolution.

\item IPCs (imaging proportional counters; e.g. ROSAT/PSPC). These are position
sensitive proportional counters (PSPC). They have rather modest energy
resolution.

\item CCDs (charge coupled devices); solid-state semiconductor detectors
developed for X-ray instruments (XMM-Newton).  They are sensitive up to
$\lambda\approx$\,100\,\AA\ and have rather modest energy resolution in the
EUV. They are different from ``usual'' CCDs (i.e. those recording optical
photons)  which were often used in UV/EUV experiments in combination with
UV--optical converters (so-called intensified CCDs, see below).

\end{itemize}

In EUV experiments, MCPs are mostly used because of the limited wavelength range
of proportional counters. Quantum efficiencies of MCPs are relatively low
($\approx$\,5\%). They can be increased significantly by depositing a
photocathode on their front face. The materials are mostly alkali-halides
(e.g. MgF$_2$, CsI, KBr). Different read-out systems for 2-d MCPs are used,
e.g., a wedge-and-strip anode (WSA), the mostly used design (e.g. EUVE). MCPs
are solar blind.

\subsubsection{EUV telescopes}

The first orbiting EUV telescopes (after Apollo-Soyuz/EUVT) were
grazing-incidence telescopes of Wolter types I and II (EUVE, ROSAT-WFC). Nested
mirrors increase the collecting area (type I Wolter). Grazing-incidence is
necessary for wavelengths below $\lambda \approx$\,300\,\AA. However, the
incidence angles can be rather large (10$^\circ$) compared to X-ray telescopes
(1$-$2$^\circ$), because the reflectivity is still quite high in EUV. Likewise,
the tolerances on surface roughness are not so severe as for X-rays. Materials
are usually glass or metal shells coated with gold.

At wavelengths larger than $\lambda\approx 500$\,\AA\ normal-incidence
telescopes with relatively high reflectivity by SiC are possible. Aluminum
coated with LiF or MgF$_2$ is preferred above $\lambda\lambda$\,1050\,\AA\ and
1150\,\AA, respectively (e.g. ORFEUS, HUT, HST, FUSE).

More recently, normal-incidence telescopes are used even at short wavelengths,
i.e., $\lambda<300$\,\AA. They use interference techniques for reflection with
multilayer coatings of optical surfaces (ALEXIS, 6 different telescopes with
large collecting area). The price to pay is a narrow bandpass. Its width is
about 10\% of the peak wavelength.

Simple mechanical collimators instead of imaging telescopes were employed in a
number of EUV missions that particularly aimed at measuring the Galactic EUV
background radiation and diffuse emission from the interstellar medium
(e.g. CHIPS, EURD), but also for stellar observations (e.g. Voyager/UVS).

\subsubsection{EUV photometry and spectroscopy}

Thin film filters and optical coating materials define bandpasses for photometry
and block unwanted background radiation. There are two categories of filters:

\begin{itemize}
\item Plastic films; material: hydrocarbon polymers (polypropylene, parylene, Lexan), and
\item Metal foils (often used with mechanical support grids).
\end{itemize}

Spectroscopic instruments in the EUV utilize reflection (EUVE, CHIPS) and
transmission gratings (Chandra/LETG, EXOSAT/TGS).  Filters can suppress
higher-order spectra.

\subsubsection{Notes on specific EUV missions}

For a long time, it was thought that observations in the EUV band would be
impossible, because of the presumed high opacity of the interstellar medium
[59All]. Encouraged by new evidence on the average density of neutral hydrogen
and probable inhomogeneities, a first successful search for EUV sources was
performed in 1975 with an instrument on the Apollo-Soyuz Test Project
[77Bow]. Further detections were made in the soft-energy range of X-ray
observatories (Einstein, EXOSAT). Important milestones were the photometric
all-sky surveys conducted with the wide-field camera (WFC) aboard ROSAT and with
the EUVE satellite.  For the first (and hitherto the only) time, EUVE offered to
a wide astronomical community access to a spectroscopic instrument covering the
entire EUV range. ALEXIS is remarkable and probably pathbreaking, because it
utilized for the first time a normal-incidence telescope in the far-EUV range.

A few UV instruments were also capable of performing near-EUV spectroscopy,
e.g., HUT, ORFEUS/BEFS, and the UVS instruments on the Voyager probes. At the
time of writing (2009), the UVS on Voyager~1 is still operating, more than 30
years after launch. It stares at a point in anti-sun direction and returns
spectra of the interplanetary medium. These are used to study the Ly$\alpha$
emission from interstellar gas penetrating the solar system [09Hol]. Currently,
there is no EUV instrument in operation, and there are no accepted proposals for
future EUV missions. The XMM-Newton and Chandra X-ray observatories provide
access to the very-far EUV ($\lambda<$\,150\,\AA), only.

\newpage

\subsection{UV instruments} 

A concise overview about the current state of technology for UV
space-instrumentation and possible future developments is given in a special
section of [99Mor].

\subsubsection{UV detectors}

In the first instruments, and up to the mid-1980s, photomultipliers were
used for photometric work.  Before the availability of 2-d detectors, spectra
were scanned with photomultipliers, too (e.g. OAO-2, Copernicus, Apollo 17,
Astron).

In some instruments (OAO-2, Copernicus, IUE) television (TV) camera tubes were
used as 2-d detectors. The secondary electron conduction (SEC) vidicon detectors
aboard IUE integrated spectrographic images on a potassium chloride SEC target
and were then read out by scanning the SEC target with an electron beam. The
photocathode of the TV camera tubes, designed for visible light response,
required a converter (with a CsTe photocathode) to transform UV into visible
radiation.

Image recording on photographic film was an important detection technique for a
long time. Two primary methods were used: one is electronography, in which
high-energy electrons from the photocathode cause the blackening of the grains
directly; the other makes use of a phosphor screen to convert the electron
energy into visible light, which is then recorded on ordinary, light-sensitive
photographic film. In the latter method the electronic imaging device serves
mainly to intensify the image, so as to increase the speed of the ordinary
photographic process, and, in some cases, to extend its spectral range into
wavelength ranges to which the film is not directly sensitive. Hence, these
latter type devices are commonly called image intensifiers or image converters
[71Car]. Film detectors were exclusively used in short-term, American and Soviet
manned  space missions, commencing with Gemini flights in 1966, and ending with
several instruments on Space Shuttles, up to 1995 (UIT, FUVIS).

Today, MCPs as well as CCDs (the latter at wavelengths longer than $\lambda
\approx 1500$\,\AA) are utilized. The UV response of CCDs is sometimes enhanced by
addition of a phosphor, such as Lumogen on the HST/WFPC-2 CCDs. CCDs with
improved UV response are under active development. Intensified CCDs (ICCD) are
image intensifier tubes or MCPs in front of optically sensitive CCDs, e.g.,
MCP-intensified CCDs in the Optical Monitor (OM) of XMM-Newton.  The first use
of an MCP detector was on the Voyager UV-Spectrographs; they served as image
intensifiers for a newly developed self-scanned linear anode array (so-called
Ssanacon) [77Bro].

Another detector type are intensified photodiode arrays. One realisation was an
MCP with a phosphor screen anode put in front of a Reticon, that is, a
visual-photon recording photodiode array (HUT). Digicons were used in the GHRS
and FOS spectrographs aboard HST. They consist of a photocathode, magnetic and
electrostatic focussing coils and electrodes, and a diode array. The
photoelectrons are accelerated directly onto the diodes with no conversion into
visible-light photons. Digicons gave way after the first-generation of HST
instruments to either multi-anode microchannel arrays (MAMAs, HST/STIS
instrument) which achieve small pixels with high photon-count rates, or
electron-bombarded CCDs.

\subsubsection{UV telescopes}

Normal-incidence mirrors are used. As for the coatings of mirrors (and optical
elements), different materials are  necessary for different wavelength bands in
order to achieve high performance. For example, the  HST mirror coating
(Al+MgF$_2$) is ``blind'' for $\lambda<1150$\,\AA, hence, other materials are
needed for high reflectivity when going to the far-UV (towards the hydrogen
Lyman edge).  For example, the FUSE instrument consists of four coaligned
telescopes and spectrographs. Two channels
with SiC coatings cover the range $\lambda\lambda$~905--1100\,\AA, and two
channels with Al+LiF coatings cover the range $\lambda\lambda$~1000--1195\,\AA\ (04Moo).

In earlier instruments refractive optics cameras were employed for imaging
(Gemini), and later on Schmidt cameras. Open collimators, i.e., non-imaging
instruments, were also used for UV-background studies since the earliest mission
(Kosmos 51) until more recently (EURD). Also, point source detection and
spectroscopy with non-imaging instruments were performed (Apollo~17, Voyager 1
\& 2, IMAPS).

\subsubsection{UV photometry, spectroscopy, and polarimetry}

In analogy to instruments working in the optical wavelength band, filters,
coating materials and detector efficiency characteristics define bandpasses for
photometry and block unwanted background radiation. Likewise, the principles of
employed spectroscopic (prisms, grisms, gratings) and polarimetric devices are
in principle the same as in optical astronomy.

\subsubsection{Notes on specific UV missions}

The early-phase history of UV space instruments culminated in 1968 with the
launch of OAO-2, the first observatory-type mission. A similar success was OAO-3
(Copernicus), which obtained first high-resolution UV spectra. TD-1A and ANS
were important UV-survey instruments. IUE was an outstanding success. For almost
two decades it  was operated as a true observatory with real-time access by
guest observers at ground stations, and delivered more than 100,000 UV spectra of all
kinds of sources.  FUSE was seminal because  it took high-resolution far-UV
spectra ($\lambda\lambda$~912--1180\,\AA) for many years. GALEX made a deep UV
all-sky survey.

The Hubble Space Telescope (HST) carries a diverse suite of imaging and
spectroscopic instruments working in a broad wavelength range spanning from the
UV to the infrared. It is the only orbiting space telescope that was frequently
maintained by Space Shuttle crews repairing instruments and installing new
ones. In 2004, HST lost its UV spectroscopic capability when the STIS
spectrograph failed. Hence, at the time of writing (2009), no UV spectroscopic
instrument is available at all (except for the small UV/optical monitors aboard
the XMM-Newton and Swift X-ray observatories). For 2009, the last of HST
servicing missions is planned, during which STIS shall be repaired and a new UV
spectrograph (COS) shall be installed. If successful, then HST will again
provide access to the UV for another $\approx$\,5 years.

There are no accepted proposals for spectroscopic UV missions beyond HST, except
for a Russian-led multi-national initiative, the World Space Observatory
Ultraviolet (WSO/UV). It shall perform imaging and spectroscopy in the
$\lambda\lambda$~1020--3200\,\AA\ band (resolving power R=1000 and 50\,000) and
is planned for launch in 2014. Two Indian-led multiple-band UV-imaging missions
(Astrosat/UVIT and GSat4/TAUVEX) are scheduled for launch in 2009.

\begin{landscape}
\begin{table}[!t]
\caption{List of space-based UV and EUV instruments. Abbreviations: ph.=photometry, im.=imaging, sp.=spectroscopy. The numbers in the
  column ``Instr.'' refer to the description of telescopes, instruments, and detectors in Table\,\ref{tabinstr}. $\Delta\lambda$ and
  R denote spectral resolution and resolving power, respectively.}
 \label{table1a}
 \begin{center}
 \begin{tabular}{llllcll}
  \hline\noalign{\smallskip}
Mission/Instrument & Country &Period&Instr.&Wavelength  & Remarks & Ref.\\
                   &or Agency&    &      &Range [\AA] &         & \\
  \hline\noalign{\smallskip}
Kosmos 51       &USSR    &1964    & 1,7   & 2300--7500 & 2 bands, UV background& 70Di1,72Dim\\   
1964 83C        &USA     &1964    & 2,7   & 1300--1650 & 1 band, stars & 67Smi\\
Venera 2, 3     &USSR    &1965    & 1,4   & 1050--1340 & 2 bands, UV background & 67Kur,68Kur\\
Gemini 10--12/S-013&USA  &1966    & 3,6,16& 2500--4400 & $\Delta\lambda$=7$-$20\,\AA, stars & 70Kon,74Spe\\
Kosmos 213      &USSR    &1968    & 1,7   & 2000--6500 & 2 bands, UV background& 70Di1,72Dim\\ 
Kosmos 215      &USSR    &1968    & 1,7   & 1250--2800 & several telescopes/bands, stars & 70Di2,76Dim\\ 
OAO-2           &USA     &1968-73 & 3,8,16& 1000--4250 & multiband im., sp.: $\Delta\lambda$=12 and 22\,\AA & 72Cod\\
Mariner 9/UVS   &USA     &1971    & 2,7,17& 1150--3400 & $\Delta\lambda$=7.5 and 15\,\AA & 71Pea,75Mol\\
Salyut\,1/Orion\,1&USSR  &1971    & 3,6,16& 2000--3800 & R=500 & 72Gur\\
STP 72-1        &USA     &1972    & 1,10  &  912--1050 & 1 band, UV background & 84Opa\\ 
Apollo\,16/S-201&USA     &1972    & 3,6,14& 1050--1600 & 2 bands, operated on lunar surface & 73Car,83Car\\
Apollo\,17/S-169&USA     &1972    & 1,7,17& 1180--1680 & $\Delta\lambda$=10\,\AA &73Fas,75Hry\\
Copernicus (OAO-3)&USA   &1972-81 & 3,7,17&  912--3275 & $\Delta\lambda$=0.05$-$0.4\,\AA &73Rog\\
TD-1A           &ESRO    &1972-74 & 3,7,17& 1350--2550 & S2/68 instr. sp. $\Delta\lambda$=35$-$40\,\AA, ph. 2750$\pm$310\,\AA & 73Bok \\
                &        &        & 3,7,17& 2000--2900 & S59 instr. $\Delta\lambda$=1.7\,\AA & 74Jag\\
Soyuz\,13/Orion\,2&USSR  &1973    & 3,6,15& 2000--4000 & $\Delta\lambda$=8-28\,\AA & 76Gur\\
Skylab 2-4      &USA     &1973-74 & 3,6,15& 1300--5000 & S-019: $\Delta\lambda$=2$-$42\,\AA & 75Hen,77Oca\\
                &        &        & 3,6,18& 2200--3000 & S-183: one band, French instrument & 77Lag,88Vui\\
                &        &        & 3,6,14& 1100--1500 & S-201: one band & 74Car\\
ANS             &NL      &1974-76 & 3,7,17& 1500--3250 & multiple bands& 75Dui,82Wes \\
D2B-Aura/ELZ    &F       &1975    & 3,7,16& 1100--3300 & multiple bands, UV background & 78Mau\\
Apollo-Soyuz/EUVT&USA,USSR&1975   & 3,10  &   40--1550 & multiple bands, 1st EUV source detections&77Bow,80Par\\
Solrad-11B      &USA     &1976    & 1,7   & 1220--1500 & 1 band, UV background & 83Wel\\
Prognoz-6/Galaktika&USSR &1977-78 & 1,7,17& 1100--1850 & $\Delta\lambda$=100\,\AA, UV background & 82Zve \\
Voyager 1+2/UVS &USA     &1977-   &1,12,16&  500--1700 & $\Delta\lambda$=18\,\AA, 1st EUV star spectra&77Bro\\
Einstein/HRI,OGS&USA     &1978-81 & 3,9,18&   3--120   & HRI: multiband im., with OGS: sp. R=10$-$50 & 79Gia \\
IUE             &Intl.   &1978-96 & 3,8,17& 1150--3300 & USA, ESA, UK; $\Delta\lambda$=0.1 and 6\,\AA &78Bog,97Har \\
Spacelab\,1/VWFC&USA     &1983    & 3,6,14& 1250--3000 & Shuttle flight, 3 bands, French instrument&94Tob \\
Astron          &USSR    &1983-89 & 3,7,17& 1100--3500 & $\Delta\lambda$=0.4 and 28\,\AA & 84Boy,86Boy\\
\noalign{\smallskip}\hline  
\end{tabular}
\end{center}
\end{table}
\end{landscape}

\addtocounter{table}{-1}
\begin{landscape}
\begin{table}[!t]
\caption{(continued)}
 \begin{center}
 \begin{tabular}{llllcll}
  \hline\noalign{\smallskip}
Mission/Instrument & Country &Period&Instr.&Wavelength  & Remarks & Ref.\\
                   &or Agency&    &      &Range [\AA] &         & \\
  \hline\noalign{\smallskip}
EXOSAT/CMA,TGS  &ESA     &1983-86 & 3,9,18  &    6--400  & CMA: multib.im., with TGS:  sp. $\Delta\lambda$=1$-$4\,\AA & 81Tay,91Whi\\
Dynamics Explorer 1&USA  &1984-87 & 1,7     & 1350--2000 & 1 band, UV background & 89Fix\\
UVX             &USA     &1986    & 3,9,17  & 1200--3100 & Shuttle flight, $\Delta\lambda$=15$-$27\,\AA, UV background&89Mur,91Hur\\
Mir/GLAZAR      &USSR    &1987-90 & 3,6,14  & 1500--1800 & 1 band & 88Tov,09Tov \\
HUT (Astro1+2)  &USA     &1990,95 &3,11,14,17& 415--1850 & 2 Shuttle flights, $\Delta\lambda$=1.5$-$3\,\AA & 92Dav,95Kru \\
UIT (Astro1+2)  &USA     &1990,95 &3,6,14,18& 1200--3400 & 2 Shuttle flights, multiband im., sp.: $\Delta\lambda$=19\,\AA  &97Ste\\
WUPPE (Astro1+2)&USA     &1990,95&3,11,14,17,20&1350--3300&2 Shuttle flights, $\Delta\lambda$=8\,\AA &94Nor \\
ROSAT/WFC,PSPC  &D,UK    &1990-99 & 3,5,9   &    5--600  & WFC\,multib.im.\,60$-$600\,\AA;PSPC\,5$-$120\,\AA,R$\approx$1&03Bar,08Pfe \\
HST             &USA,ESA &1990-   & see text& 1150--IR   & several instruments, see text&09Hst \\
FUVCam          &USA     &1991    & 3,6,14  & 1220--2000 & Shuttle flight, 2 bands &95Sch \\
FAUST           &USA     &1992    & 3,9     & 1400--1800 & Shuttle flight, 1 band & 93Bow\\
DUVE            &USA     &1992    & 1,9,17  &  950--1080 & $\Delta\lambda$=3\,\AA, diffuse emission from hot ISM & 98Kor\\
EUVE            &USA     &1992-01 & 2,9,17  &   60--800  & multiband im., sp.: R=200&91Boa \\
ALEXIS          &USA     &1993-05 & 3,9     &  130--186  & 3 narrow bands, normal incidence optics &96Blo\\
ORFEUS          &D,USA   &1993,96 & 3,9,17  &  380--1400 & 2 Shuttle flights, BEFS: 380$-$1220\,\AA, R=4600 & 96Hur\\ 
                &        &        &         &            & TUES: 912-1400\,\AA, R=11\,000 &99Bar\\
IMAPS           &USA     &1993,96 &1,13,14,16& 950--1150 & 2 Shuttle flights with ORFEUS, R=120\,000 & 96Jen\\
FUVIS           &USA     &1995    &3,6,14,17?& 970--2000 & Shuttle flight, $\Delta\lambda$=5 and 30\,\AA, diffuse sources & 90Car\\
UVSTAR      &Italy,USA   &1995-98 &3,13,14,17& 535--1250 & 3 Shuttle flights (1995,97,98), $\Delta\lambda$=1$-$12\,\AA & 93Sta,02Gre \\
MSX/UVISI       &USA     &1996-97 & 3,13,14 & 1240--8270 & multiple bands, primarily military mission&01Mur,04New \\
Minisat-01/EURD&Spain,USA&1997-01 & 1,9,16  &  350--1100 & $\Delta\lambda$=5\,\AA, spectral im. & 97Bow,01Mor\\ 
FUSE            &USA     &1999-07 & 3,9,17  &  912--1180 & R=20\,000 & 00Moo,00Sah\\
Chandra/LETG    &USA     &1999-   &3,9,13,18&   6--150  & R=100$-$1000 & 02Wei\\
XMM/OM,pn-CCD   &ESA     &1999-   &3,13,14,19&1700--6000 & OM: multiband im., sp. R=250 &01Mas\\
                &        &        &  3,13   &    1--100  & pn-CCD: R=1&01Str\\
CHIPS           &USA     &2003-08 & 1,9,17  &   90--260  & R=50$-$150, EUV background &03Hur,05Hur\\
GALEX           &USA     &2003-   & 3,9,19  & 1350--2800 & 2 bands im., grism sp.: R=90$-$200 &05Mar,05Mor\\
SPEAR (=FIMS)  &Korea,USA&2003-05 & 2,9,17  &  900--1750 & R=550, spectral im. & 06Ede\\
Swift/UVOT      &USA     &2004-   &3,13,14,19&1700--6500 & multiband im., sp.: R=150 &04Geh\\
\noalign{\smallskip}\hline  
\end{tabular}
\end{center}
\end{table}
\end{landscape}

\begin{table}[!t]
\caption{Instrumentation reference for UV and EUV space missions listed in Table~\ref{table1a}.}
 \label{tabinstr}
 \begin{center}
 \begin{tabular}{lll}
  \hline\noalign{\smallskip}
  & \# & Description \\
  \hline\noalign{\smallskip}
Telescope type    &  1 & mechanical collimator \\
                  &  2 & collecting mirror \\
                  &  3 & imaging mirror \\
\noalign{\smallskip}
Detector          &  4 & Geiger counter \\
                  &  5 & gas proportional counter \\
                  &  6 & photographic film \\
                  &  7 & photomultiplier \\
                  &  8 & television tube (including UV--visible photon converter) \\
                  &  9 & microchannel plate (MCP) \\ 
                  & 10 & channel electron multiplier (CEM) \\
                  & 11 & photodiode detector (Reticon, Digicon) \\
                  & 12 & MCP-intensified self-scanned linear anode array (Ssanacon) \\
                  & 13 & charge-coupled device (CCD) \\
                  & 14 & image intensifier (e.g. MCP); acts as UV--visible photon converter \\
                  &    & in front of optical photon detectors (film, CCDs, photodiodes)\\
\noalign{\smallskip}
Dispersive element& 15 & objective prism \\
                  & 16 & objective grating \\
                  & 17 & reflection grating \\
                  & 18 & transmission grating \\
                  & 19 & grism \\
\noalign{\smallskip}
Polarimetry       & 20 & polarizing beamsplitter\\
\noalign{\smallskip}\hline  
\end{tabular}
\end{center}
\end{table}

\subsection{References for~\thesection}
\vspace{-0.5cm}
\noindent
\begin{longtable}{ @{}p{1.0cm}@{}p{14.0cm}@{} }
59All & Aller, L.H.: Publ. Astron. Soc. Pacific \textbf{71} (1959) 324.\\
67Smi & Smith, A.M.: Astrophys. J. \textbf{147} (1967) 158.\\
67Kur & Kurt, V.G., Sunyaev, R.A.: Soviet J. Exp. Theoret. Ph. Letters \textbf{5} (1967) 246.\\
68Kur & Kurt, V.G., Sunyaev, R.A.: Soviet Astron. \textbf{11} (1968) 928.\\
69Wil & Wilson, R., Boksenberg, A.: Annu. Rev. Astron. Astrophys. \textbf{7} (1969) 421.\\
70Di1 & Dimov, N.A., Severny, A.B., Zvereva, A.M.: in Ultraviolet Stellar Spectra and Ground-Based Observations, eds. L. Houziaux,
Butler, H.-E., IAU Symp. \textbf{36} (1970), 325.\\
70Di2 & Dimov, N.A.: in Ultraviolet Stellar Spectra and Ground-Based Observations, eds. L. Houziaux,
Butler, H.-E., IAU Symp. \textbf{36} (1970), 138.\\
70Kon & Kondo, Y., Henize, K.G., Kotila, C.L.: Astrophys. J. \textbf{159} (1970) 927.\\
71Car & Carruthers, G.R.: Astrophys. Sp. Science \textbf{14} (1971) 332.\\
71Pea & Pearce, J.B., Gause, K.A., Mackey, E.F., Kelly, K.K., Fastie, W.G., Barth, C.A.: Appl. Optics \textbf{10} (1971) 805.\\
72Ble & Bless, R.C., Code, A.D.:  Annu. Rev. Astron. Astrophys. \textbf{10} (1972) 197.\\
72Cod & Code, A.D. (ed.).: The Scientific Results from the Orbiting Astronomical Observatory (OAO-2), NASA SP-310 (1972).\\
72Dim & Dimov, N.A., Zvereva, A.M., Severny, A.B.: Izv. Krymskoj Astrofiz. Obs. \textbf{45} (1972) 67.\\
72Gur & Gurzadyan, G.A., Ohanesyan, J.B.: Space Sci. Rev. \textbf{13} (1972) 647.\\
73Bok & Boksenberg, A., Evans, R.G., Fowler, R.G., et al.: Mon. Not. R. Astr. Soc. \textbf{163} (1973) 291.\\
73Car & Carruthers, G.R.: Appl. Opt. \textbf{12} (1973) 2501.\\
73Fas & Fastie, W.G.: The Moon \textbf{7} (1973) 49.\\
73Rog & Rogerson, J.B., Spitzer, L., Drake, J.F., et al.:  Astrophys. J. Lett. \textbf{181} (1973) L97.\\
74Car & Carruthers, G.R., Opal, C.B., Page, T.L., Meier, R.R., Prinz, D.K.: Icarus \textbf{23} (1974) 526.\\
74Jag & de Jager, C., Hoekstra, R., van der Hucht, K.A.: Astrophys. Sp. Science \textbf{26} (1974) 207.\\
74Spe & Spear, G.G., Kondo, Y., Henize, K.G.: Astrophys. J. \textbf{192} (1974) 615.\\
75Dui & van Duinen, R.J., Aalders, J.W.G., Wesselius, P.R., Wildeman, K.J., Wu, C.C., Luinge, W., Snel, D.: Astron. Astrophys.
         \textbf{39} (1975) 159.\\
75Hen & Henize, K.G., Wray, J.D., Parsons, S.B., et al.: Astrophys. J. Lett. \textbf{199} (1975) L119.\\
75Hry & Henry, R.C., Weinstein, A., Feldman, P.D., Fastie, W.G., Moss, H.W.:  Astrophys. J. \textbf{201} (1975) 613.\\
75Mol & Molnar, M.R.: Astrophys. J. \textbf{200} (1975) 106.\\
76Dim & Dimov, N.A., Teres, E.I.: Izv. Krymskoj Astrofiz. Obs. \textbf{55} (1976) 196.\\
76Gur & Gurzadyan, G.A., Jarakyan, A.L., Krmoyan, M.N., Kashin, A.L., Loretsyan, G.M., Ohanesyan, J.B.: Space Sci. Rev.
         \textbf{40} (1976) 393.\\
77Bow & Bowyer, S., Margon, B., Lampton, M., Paresce, F., Stern, R.: Apollo-Soyuz Test Project Summary Science Report (Vol. 1), NASA
          SP-412 (1977) p. 49.\\
77Bro & Broadfoot, A.L., Sandel, B.R., Shemansky, D.E., et al.: Space Sci. Rev. \textbf{21} (1977) 183.\\
77Lag & Laget, M., Saisse, M., Vuillemin, A.: Appl. Opt. \textbf{16} (1977) 961.\\ 
77Oca & Ocallaghan, F.G., Henize, K.G., Wray, J.D.: Appl. Opt. \textbf{16} (1977) 973.\\
78Bog & Boggess, A., Carr, F.A., Evans, D.C., et al.: Nature \textbf{275} (1978), 372.\\
78Mau & Maucherat-Joubert, M., Cruvellier, P., Deharveng, J.M.: Astron. Astrophys. \textbf{70} (1978) 467.\\
79Gia & Giacconi, R., Branduardi, G., Briel, U., et al.: Astrophys. J. \textbf{230} (1979) 540.\\
80Par & Paresce, F., McKee, C., Bowyer, S.: Astrophys. J. \textbf{240} (1980) 387.\\
81Tay & Taylor, B.G., Andresen, R.D., Peacock, A., Zobl, R.: Space Sci. Rev. \textbf{30} (1981) 479.\\
82Wes & Wesselius, P.R., van Duinen, R.J., de Jonge, A.R.W., et al.: Astron. Astrophys. Suppl. Ser. \textbf{49} (1982) 427.\\
82Zve & Zvereva, A.M., Severny, A.B., Granitzky, L.V., et al.: Astron. Astrophys. \textbf{116} (1982) 312.\\
83Car & Carruthers, G.R., Page, T.: Astrophys. J. Suppl. \textbf{53} (1983) 623.\\
83Wel & Weller, C.S.: Astrophys. J. \textbf{268} (1983) 899.\\
84Boy & Boyarchuk, A.A., Gershberg, R.E., Granitskij, L.V., et al.:  Sov. Astron. Lett. \textbf{10} (1984) 67.\\
84Opa & Opal, C.B., Weller, C.: Astrophys. J. \textbf{282} (1984) 445.\\
86Boy & Boyarchuk, A.A.: Irish Astron. J. \textbf{17} (1986) 352.\\
88Tov & Tovmasyan, G.M., Khodzhayants, Yu.M., Krmoyan M.N., et al.: Sov. Astron. Lett., \textbf{14} (1988) 123.\\
88Vui & Vuillemin, A.: Astron. Astrophys. Suppl. Ser. \textbf{72} (1988) 249.\\
89Fix & Fix, J.D., Craven, J.D., Frank, L.A.: Astrophys. J. \textbf{345} (1989) 203.\\
89Mur & Murthy, J., Henry, R.C., Feldman, P.D., Tennyson, P.D.: Astrophys. J. \textbf{336} (1989) 954.\\
90Car & Carruthers, G.R., Heckathorn, H.M., Witt, A.N., Raymond, J.C., Opal, 
        C.B., Dufour, R.J.: in The Galactic and Extragalactic Background Radiation, IAU
        Symp. 139, Reidel, Dordrecht (1990) 459.\\
91Boa & Bowyer, S., Malina, R.F.: Extreme Ultraviolet Astronomy, eds. R.F. Malina, S. Bowyer, Pergamon Press, Elmsford (1991) 397.\\
91Bob & Bowyer, S.: Annu. Rev. Astron. Astrophys. \textbf{29} (1991) 59.\\
91Hur & Hurwitz, M., Bowyer, S., Martin, C.: Astrophys. J. \textbf{372} (1991) 167.\\
91Mal & Malina, R.F., Bowyer, S. (eds.): Extreme Ultraviolet Astronomy, Pergamon Press (1991).\\
91Whi & White, N.E.: Extreme Ultraviolet Astronomy, eds. R.F. Malina, S. Bowyer, Pergamon Press (1991) 15.\\
92Dav & Davidsen, A.F., Long, K.S., Durrance, S.T., et al.: Astrophys. J. \textbf{392} (1992) 264.\\
93Bow & Bowyer, S., Sasseen, T.P., Lampton, M., Wu, X.: Astrophys. J. \textbf{415} (1993) 875.\\
93Sta & Stalio, R., Sandel, B.R., Broadfoot, A.L., Chavez, M.: Adv. Space Res. \textbf{13} (1993) 379.\\
94Nor & Nordsieck, K.H., Code, A.D., Anderson, C.M.: X-ray and Ultraviolet Polarimetry, ed. S. Fineschi, Proc. SPIE, \textbf{2010}
(1994) 2.\\
94Tob & Tobin, W., Viton, M., Sivan, J.-P.: Astron. Astrophys. Suppl. Ser. \textbf{107} (1994) 385.\\
95Kru & Kruk, J.W., Durrance, S.T., Kriss, G.A., et al.: Astrophys. J. \textbf{454} (1995) L1.\\
95Sch & Schmidt, E.G., Carruthers, G.R.: Astrophys. J. Suppl. \textbf{96} (1995) 605.\\
96Blo & Bloch, J.J.: Astrophysics in the Extreme Ultraviolet, eds. S. Bowyer, R.F. Malina, Kluwer, Dordrecht (1996), 7.\\
96Bow & Bowyer, S., Malina, R.F. (eds.): Astrophysics in the Extreme Ultraviolet, IAU Coll. 152, Kluwer, Dordrecht (1996).\\
96Hur & Hurwitz, M., Bowyer, S.: Astrophysics in the Extreme Ultraviolet, eds. S. Bowyer, R.F. Malina, Kluwer, Dordrecht (1996), 601.\\
96Jen & Jenkins, E.B., Reale, M.A., Zuccino, P.M., Sofia, U.J.: Astrophys. Space Sci. \textbf{239} (1996), 315.\\
97Bow & Bowyer, S., Edelstein, J., Lampton, M.:  Astrophys. J. \textbf{485} (1997) 523.\\
97Har & Harris, R.A. (ed.): IUE Spacecraft Operations, Final Report, ESA SP-1215 (1997).\\
97Ste & Stecher, T.P., Cornett, R.H., Greason, M.R.: Publ. Astron. Soc. Pacific \textbf{109} (1997) 584.\\
98Kor & Korpela, E.J., Bowyer, S., Edelstein, J.: Astrophys. J. \textbf{495} (1998) 317.\\
99Bar & Barnstedt, J., Kappelmann, N., Appenzeller, I., et al.: Astron. Astrophys. Suppl. Ser. \textbf{134} (1999) 561.\\
99Mor & Morse, J.A., Shull, J.M., Kinney, A.L. (eds.): Ultraviolet-Optical Space Astronomy Beyond HST, ASP Conf. Ser. \textbf{164} (1999).\\
00Bow & Bowyer, S., Drake, J.J., Vennes, S.: Annu. Rev. Astron. Astrophys. \textbf{38} (2000) 231.\\ 
00Moo & Moos, H.W., Cash, W.C., Cowie, L.L., et al.: Astrophys. J. \textbf{538} (2000) L1.\\
00Sah & Sahnow, D.J., Moos, H.W., Ake, T.B., et al.: Astrophys. J. \textbf{538} (2000) L7.\\
01Mas & Mason, K.O., Breeveld, A., Much, R., et al.: Astron. Astrophys. \textbf{365} (2001) L36.\\
01Mor & Morales, C., Orozco, V., G\'omez, J.F.:  Astrophys. J. \textbf{552} (2001) 552.\\
01Mur & Murthy, J., Henry, R.C., Paxton, L.J., Proce, S.D.: Bull. Astr. Soc. India \textbf{29} (2001) 563.\\
01Str & Str\"uder, L., Briel, U., Dennerl, K.:  Astron. Astrophys. \textbf{365} (2001) L18.\\
02Gre & Gregorio, A., Stalio, R., Broadfoot, L., Castelli, F., Hack, M., Holberg, J.: Astron. Astrophys. \textbf{383} (2002) 881.\\
02Wei & Weisskopf, M.C., Brinkman, B., Canizares, C., et al.: Publ. Astron. Soc. Pacific \textbf{114} (2002) 1.\\
03Bar & Barstow, M.A., Holberg, J.B.: Extreme Ultraviolet Astronomy, Cambridge University Press (2003).\\
03Hur & Hurwitz, M.: Proc. SPIE \textbf{5164} (2003) 24.\\
04Geh & Gehrels, N., Chincarini, G., Giommi, P., et al.: Astrophys. J. \textbf{611} (2004) 1005.\\
04Moo & Moos, H.W., McCandliss, S.R., Kruk, J.W.: SPIE \textbf{5488} (2004) 1.\\
04New & Newcomer, R.E., Murthy, J., Henry, R.C., Price, S.D., Paxton, L.: Air Force Research Laboratory Technical Report
AFRL-VS-TR-2004-1056 (2004). Available as VizieR On-line Data Catalog II/269: MSX Ultraviolet Point Source Catalog. \\
05Hur & Hurwitz, M., Sasseen, T.P., Sirk, M.M.: Astrophys. J. \textbf{623} (2005) 911.\\ 
05Mar & Martin, D.C., Fanson, J., Schiminovich, D.: Astrophys. J. \textbf{619} (2005) L1.\\
05Mor & Morissey, P., Schiminovich, D., Barlow, T.A., et al.: Astrophys. J. \textbf{619} (2005) L7.\\
06Ede & Edelstein, J., Korpela, E.J., Adolfo, J., et al.: Astrophys. J. \textbf{644} (2006) L159.\\
07Val & Valcu, B., Smith, P.L., Gardner, L.D., Raymond, J.C., Miralles, M.-P., Kohl, J.L.: Solar Phys. \textbf{243} (2007) 93.\\
08Pfe & Pfeffermann, E.: The Universe in X-Rays, eds. J.E. Tr\"umper, G. Hasinger, Springer (2008) 21.\\
09Tov & Tovmassian, H.M.: priv. comm. (2009).\\
09Hol & Holberg, J.B.: priv. comm. (2009).\\
09Hst & HST webpage {\tt http://www.stsci.edu/hst/HST\_overview/} at Space Telescope Science Institute (2009).\\

\end{longtable}

\end{document}